\documentclass[aps,prd,twocolumn,showpacs, superscriptaddress,amsmath,amssymb,nofootinbib]{revtex4-1}

\usepackage[pdftex]{graphicx}
\DeclareGraphicsExtensions{.pdf}
\usepackage{wasysym}
\usepackage{color}

\begin{document}

\title{Nuclear recoil energy scale in liquid xenon with application to the direct detection of dark matter}

\author{Peter~Sorensen}
\email{pfs@llnl.gov}
\affiliation{Lawrence Livermore National Laboratory, 7000 East Ave., Livermore, CA 94550, USA}
\author{Carl Eric~Dahl}
\affiliation{Enrico Fermi Institute, KICP and Department of Physics, University of Chicago, Chicago, USA}

\date{\today}
\begin{abstract}
We show for the first time that the quenching of electronic excitation from nuclear recoils in liquid xenon is well-described by Lindhard theory, if the nuclear recoil energy is reconstructed using the combined (scintillation and ionization) energy scale proposed by Shutt {\it et al.}.  We argue for the adoption of this perspective in favor of the existing preference for reconstructing nuclear recoil energy solely from primary scintillation.  We show that signal partitioning into scintillation and ionization is well-described by the Thomas-Imel box model.  We discuss the implications for liquid xenon detectors aimed at the direct detection of dark matter.
\end{abstract}

\maketitle

\section{Introduction} \label{sec1}
There is considerable experimental effort dedicated to the direct detection of particle dark matter.  Among the various detection strategies \cite{2004gaitskell}, dual-phase liquid xenon detectors have recently achieved a rapid scale-up in target mass \cite{2010aprile}, which directly improves the sensitivity of the dark matter search.  The expected experimental signature from the scattering of a dark matter particle is in most cases a low-energy $\mathcal{O}$(keV) recoiling nucleus.  A fraction $f_n$ of the nuclear recoil energy E$_{nr}$ is transferred to measurable electronic excitation, and the rest is lost to heat, e.g. atomic motion.  The theoretical prediction for the quenching $f_n$ obtained by Lindhard {\it et al.} \cite{1963lindhard} agrees to within $\pm20\%$ with the measured amount of ionization in germanium detectors \cite{2007benoit,2007barbeau}.  However, it has been observed that the Lindhard theory does not agree with measurements of ionization quenching of nuclear recoils in liquid xenon \cite{2006aprile}.

Experiments with liquid xenon as the target medium historically use only the primary scintillation signal to reconstruct E$_{nr}$ \cite{2010aprile,2009angle,2009lebedenko,2008angle,2007alner}.  Commensurate with this non-linear energy scale determined by primary scintillation, there has been a substantial experimental effort aimed at measuring (see \cite{2010manzur,2009aprile} and references therein) and understanding \cite{2010tretyak,2008mei,2007mangiarotti,2005hitachi} the scintillation quenching $\mathcal{L}_{eff}$ of liquid xenon for nuclear recoils.     In spite of these efforts, factor of two disagreement persists between recent measurements, and Lindhard theory does not correctly predict $\mathcal{L}_{eff}$ \cite{2006chepel,2005aprile,2006aprile,2010tretyak,2007mangiarotti}.  However, with semi-empirical modifications, reasonable agreement may be obtained \cite{2010manzur,2008mei,2005hitachi}.  

Meanwhile, it has been known for nearly a decade \cite{2002doke} that a linear, drift-field independent energy scale with a substantial improvement in energy resolution \cite{2007aprile,2003conti} is obtained from a simultaneous measurement of the number of primary scintillation photons $n_{\gamma}$ and ionized electrons $n_{e}$.  This is given for electron and nuclear recoils by Shutt {\it et al.} \cite{2007shutt} as
\begin{equation} \label{eq1}
\mbox{E}_{er}=\epsilon(n_{\gamma} + n_{e}),
\end{equation} 
\begin{equation} \label{eq1a}
\mbox{E}_{nr}=\epsilon(n_{\gamma} + n_{e})/f_n,
\end{equation} 
where $\epsilon=13.8\pm0.9$~eV \cite{2002doke} is the average energy required to create a single quanta, either photon or electron.  Other measurements have found $\epsilon=13.7\pm0.2$~eV \cite{2009dahl} and $\epsilon=14$~eV \cite{2010aprile2}.  Eq. \ref{eq1} is widely used to measure E$_{er}$ for electron recoils from gamma and beta particles \cite{2010aprile2,2009lebedenko,2007aprile}, but Eq. \ref{eq1a} as a measure of E$_{nr}$ has been largely ignored.  We note that this formulation is analogous to the ionization energy scale in germanium detectors.
 
In this article, we show that the Lindhard theory is consistent with the quenching of the total ($n_e+n_{\gamma}$) electronic excitation from nuclear recoils in liquid xenon.  This further motivates the adoption of the combined energy scale (Eq. \ref{eq1a}), but it does not resolve the disagreement between recent measurements of $\mathcal{L}_{eff}$ \cite{2010manzur,2009aprile}.  We therefore pinpoint the likely source of the present disagreement, and suggest a strategy for reducing systematic effects in future measurements of either $\mathcal{L}_{eff}$ or $f_n$. 

Our motivation for this work is to obtain the best reconstruction of the detected nuclear recoil energy in liquid xenon dark matter search experiments.  The predicted nuclear recoil energy spectrum of halo-bound particle dark matter elastically scattering from a xenon target falls sharply with increasing E$_{nr}$ \cite{1996lewin}.  This implies that the sensitivity of a given exposure (detector mass $\times$ time) of a liquid xenon detector has a strong dependence on the detector energy scale.  For light ($\lesssim10$~GeV) dark matter particles, the dependence on energy scaling and threshold is even more severe.  Recent results from XENON100  \cite{2010aprile}, in which no candidate events were observed, gave a clear example of this.  Depending on assumptions about the energy scaling (obtained therein from primary scintillation via $\mathcal{L}_{eff}$) and the energy resolution, XENON100 may or may not exclude parameter space consistent with potential signal in the CoGeNT \cite{2010aalseth} and DAMA \cite{2008bernabei} experiments.  

\section{Reconstructing nuclear recoil energy}\label{sec2}
\begin{figure}[ht]
\centering
\includegraphics[width=0.48\textwidth]{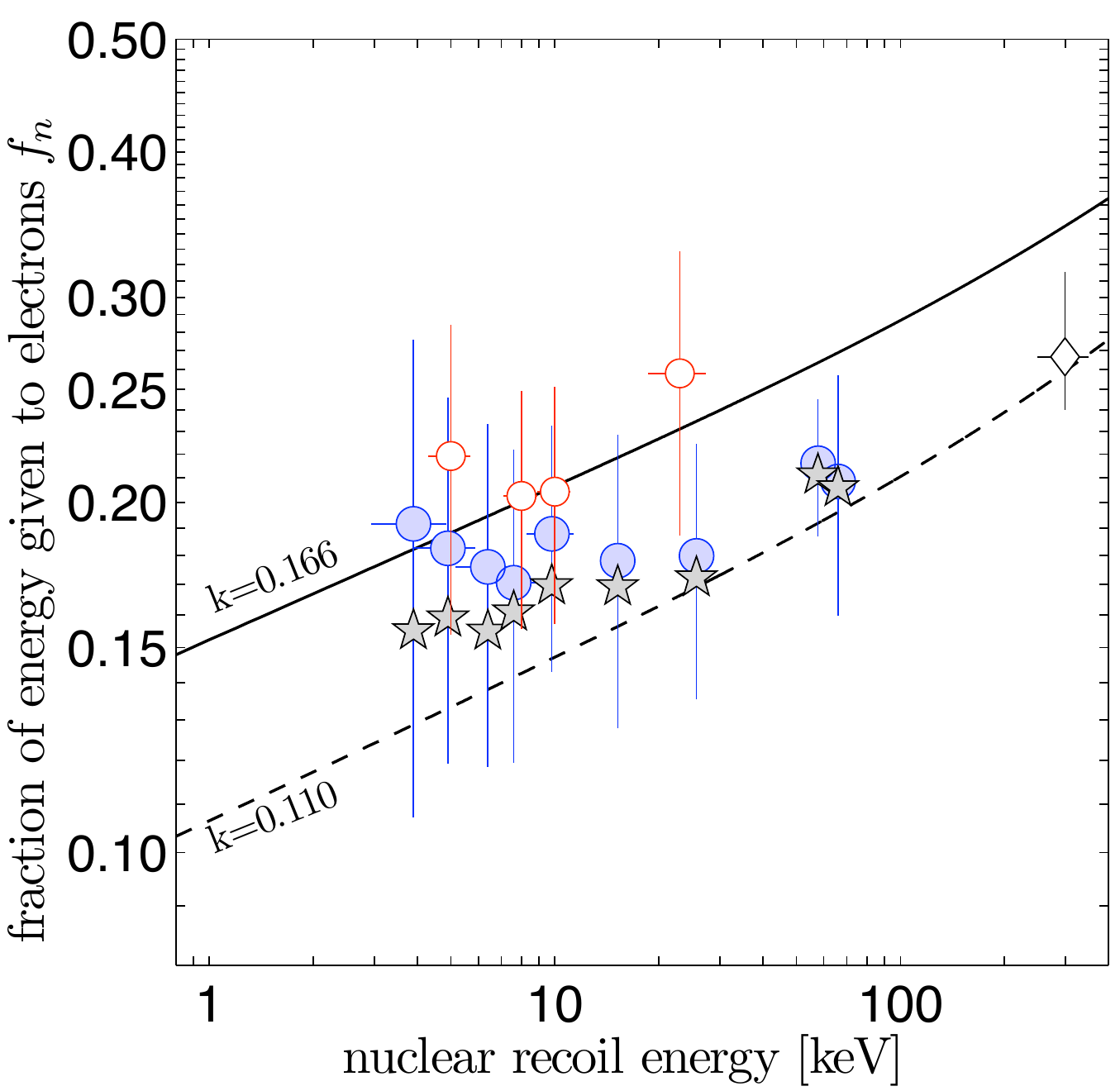}
\caption{Quenching of electronic excitation from nuclear recoils in liquid xenon: from \cite{2010manzur} ($\newmoon$), from \cite{2010manzur} as corrected by \cite{2010sorensen} ($\bigstar$, uncertainty not shown but similar to $\newmoon$), from \cite{2009aprile} ($\fullmoon$) and from \cite{2009sorensen} ($\lozenge$).  Also shown are the theoretical prediction \cite{1963lindhard} for two calculated values of $k$ (solid and dashed curves). }
\label{fig1}
\end{figure}

Lindhard {\it et al.} \cite{1963lindhard} calculated a general expression for the expected fraction of nuclear recoil energy that is transferred to electrons.  It can be written as
\begin{equation} \label{eq2}
f_n = k\cdot g / (1+k \cdot g),
\end{equation}
with $k=0.133~Z^{2/3} A^{-1/2}$.  Physically, $k$ is a proportionality constant between the electronic stopping power $d\mbox{E}/dx$ and the velocity of the projectile (which in this context is a recoiling xenon atom).  The relation is most simply expressed in terms of dimensionless variables, as in \cite{1963lindhard}.  For xenon, Lindhard's calculation results in $k=0.166$.  Recently, Hitachi calculated from first principles the electronic stopping power of recoiling xenon atoms in a liquid xenon target.  The result is shown in Fig. 5 of \cite{2006aprile}, and discussed further in \cite{2005hitachi}.  In terms of the dimensionless variables of \cite{1963lindhard}, his calculation corresponds to $k=0.110$.  Note that no analytic form was given for the energy-dependent function $g$ in \cite{1963lindhard}, and we have used the parameterization given in \cite{1996lewin}.  In Fig. \ref{fig1} we show $f_n$ as calculated from Eq. \ref{eq2}, for these two values of $k$ (solid and dashed curves).   In \cite{1963lindhard}, Lindhard {\it et al.} cautioned that ``Maybe the greatest uncertainty is the proportionality factor, k... [which] is often on the interval $0.10<k<0.20$.''  Ideally, the remedy for this uncertainty may be obtained from data.

In order to compare with data, we write Eq. \ref{eq1a} as
\begin{equation} \label{eq3}
f_n = \epsilon(\frac{\mbox{S1}}{\alpha_1} + \frac{\mbox{S2}}{\alpha_2})/\mbox{E}_{nr},
\end{equation}
with $n_{\gamma}$ and $n_e$ in terms of the experimentally measured quantities S1 and S2.  These are just the number of recorded photoelectrons in the primary scintillation and ionization (measured from proportional scintillation) signals in a dual-phase xenon detector.  The number of primary scintillation photons is $n_{\gamma}=\mbox{S1}/\alpha_1$, where $\alpha_1\sim\mathcal{O}(0.1)$ is the total efficiency to convert a scintillation photon to a detectable photoelectron.  The number of ionized electrons is $n_e=\mbox{S2}/\alpha_2$, where $\alpha_2\sim \mathcal {O}(10)$ is the number of photoelectrons registered from the proportional scintillation resulting from a single ionized electron.  While $\alpha_2$ is reasonably easy to measure in dual-phase liquid xenon detectors, $\alpha_1$ is difficult to measure directly. As a result, experiments instead quote the scintillation light yield $L_y$ (units of photoelectrons / keV) of a mono-energetic gamma source.  The proportionality constant between $\alpha_1$ and $L_y$ depends strongly on both the incident gamma energy and the electric field ($E_d$) applied across the liquid xenon target.  For a 122~keV gamma from $^{57}$Co at $E_d=0$, a detector-independent expression for the proportionality is $\alpha_1=0.015L_y$ \cite{2009dahl}.

We need $\alpha_1$ in order to use Eq. \ref{eq3}, and first cross-check the relationship given above, which implies $\alpha_1=0.080$ \cite{2009dahl} for the XENON10 detector \cite{2010aprile2}.  XENON10 measured $\alpha_2=24\pm1$ \cite{2009sorensen} from the background distribution of single electrons.  The value of $\alpha_1$ is also uniquely determined by requiring Eq. \ref{eq1} to reproduce the correct peak positions of gamma lines.  Since the energy scale is linear, any mono-energetic source will suffice.  From the 164~keV gamma observed by XENON10, we find $\alpha_1=0.078\pm0.005$.


Simultaneous measurements of the scintillation ($\mathcal{L}_{eff}$) and ionization ($\mathcal{Q}_y$) yield of liquid xenon as a function of E$_{nr}$ were obtained by Manzur {\it et al.} \cite{2010manzur}.  Their ionization yield data was presented in terms of $n_e$ (so we do not need $\alpha_2$), and their scintillation data in terms of $L_y$ for 122~keV gammas.  Using the scaling relation given above, we infer $\alpha_1$ for their detector.  We then use Eq. \ref{eq3} to cast the results from \cite{2010manzur} in terms of $f_n$.  This is shown in Fig. \ref{fig1} (as $\newmoon$, with $1\sigma$ uncertainty).   The combined $\mathcal{L}_{eff}$ and $\mathcal{Q}_y$ measurements of Manzur {\it et al.} are not quite consistent with the XENON10 nuclear recoil band measurement \cite{2009angle}.  In \cite{2010sorensen} it is argued that for the three data points below E$_{nr}\simeq6$~keV, the most likely origin of the disagreement is that $\mathcal{Q}_y$ was overstated by about $1\sigma$ due to spurious threshold effects.  The Manzur {\it et al.} data as corrected by \cite{2010sorensen} is also shown in Fig. \ref{fig1} (as $\bigstar$, uncertainty similar but omitted for clarity).  

The experiments described in \cite{2010aprile2} and \cite{2010manzur} (and, for that matter, \cite{2010aprile}) obtained their data with different values of $E_d$.  Although the values of $E_d$ ranged from about $0.5$~kV/cm \cite{2010aprile} to $1.0$~kV/cm \cite{2010manzur}, the effects of this difference are negligible.  As shown in Fig. 3 of \cite{2006aprile}, the scintillation and ionization signals from nuclear recoils exhibit no significant dependence on $E_d$, in the range $0.5-1.0$~kV/cm.  At values of $E_d\lesssim0.2$~kV/cm, a decrease of about 8\% is observed in the ionization signal, relative to $E_d\gtrsim0.5$~kV/cm.  In the same range, the scintillation signal is observed to increase by a similar amount.

Aprile {\it et al.} recently measured $\mathcal{L}_{eff}$ at $E_d=0$ \cite{2009aprile}.  This condition prevented them from making a simultaneous measurement of the ionization yield, as in \cite{2010manzur}.  Following \cite{2010sorensen}, we used the constraint from the XENON10 nuclear recoil band to infer the ionization yield (at $E_d=0.73$~kV/cm) for this data, taking care to account for the 8\% shift in $n_e$ and $n_{\gamma}$.  The resulting $f_n$ is shown in Fig. \ref{fig1} (as $\fullmoon$, with $1\sigma$ uncertainty).  Uncertainty in the ionization yield was assumed to be 20\%, which is similar to the uncertainty reported in \cite{2010manzur}.  The slightly smaller total uncertainty in this case results from the higher $\alpha_1$ obtained in that experiment. 

A final data point in Fig. \ref{fig1} ($\lozenge$) was obtained from the endpoint of the XENON10 neutron calibration \cite{2009sorensen}.  The maximum recoil energy imparted by Am-Be neutrons to a xenon target via elastic scattering is E$_{nr}=4\mbox{E}_nm_nm/(m_n+m)^2$.  The maximum incident neutron energy was  E$_n=10.5\pm0.5$~MeV \cite{1995marsh}, with the uncertainty represented by the horizontal error bar.

\section{Discussion} \label{sec3}
Signal quenching from nuclear recoils in liquid xenon appears well-described by the Lindhard model of quenching, once account is taken of all energy transferred to electrons.  This is a very significant result, but it should not be surprising.  Lindhard {\it et al.} defined $\bar{\eta}$ as ``the sum total of the energy given to electrons,'' \cite{1963lindhard} with explicit mention of both excitation and ionization.  In liquid xenon, the former leads to $n_{\gamma}$ and the latter to both $n_e$ and $n_{\gamma}$ \cite{2002doke}.  In our notation, which follows that of other recent work, $f_n \equiv \bar{\eta}/\mbox{E}_{nr}$.  

Considering the uncertainties in the existing data, it is unclear which calculated value of $k$ is preferred.  But it is clear that systematic uncertainty in future measurements of $f_n$ can be reduced.  Essential steps not taken in previous work include direct {\it in-situ} measurement of $\alpha_1$ and $\alpha_2$.  A particularly apt choice for measuring $\alpha_1$ would be the internal, homogenous 40 keV line from $^{83m}$Kr \cite{2010manalaysay,2010kastens}.  Non-uniformities in signal collection at the detector edges could be minimized by $x-y$ position reconstruction, and rejection of edge events.  As pointed out in \cite{2010sorensen}, it is also desirable to obtain a measurement of the nuclear recoil band, as a cross-check.  However, this may be difficult to implement without shielding from external background radiation.

\begin{figure}[ht]
\centering
\includegraphics[width=0.48\textwidth]{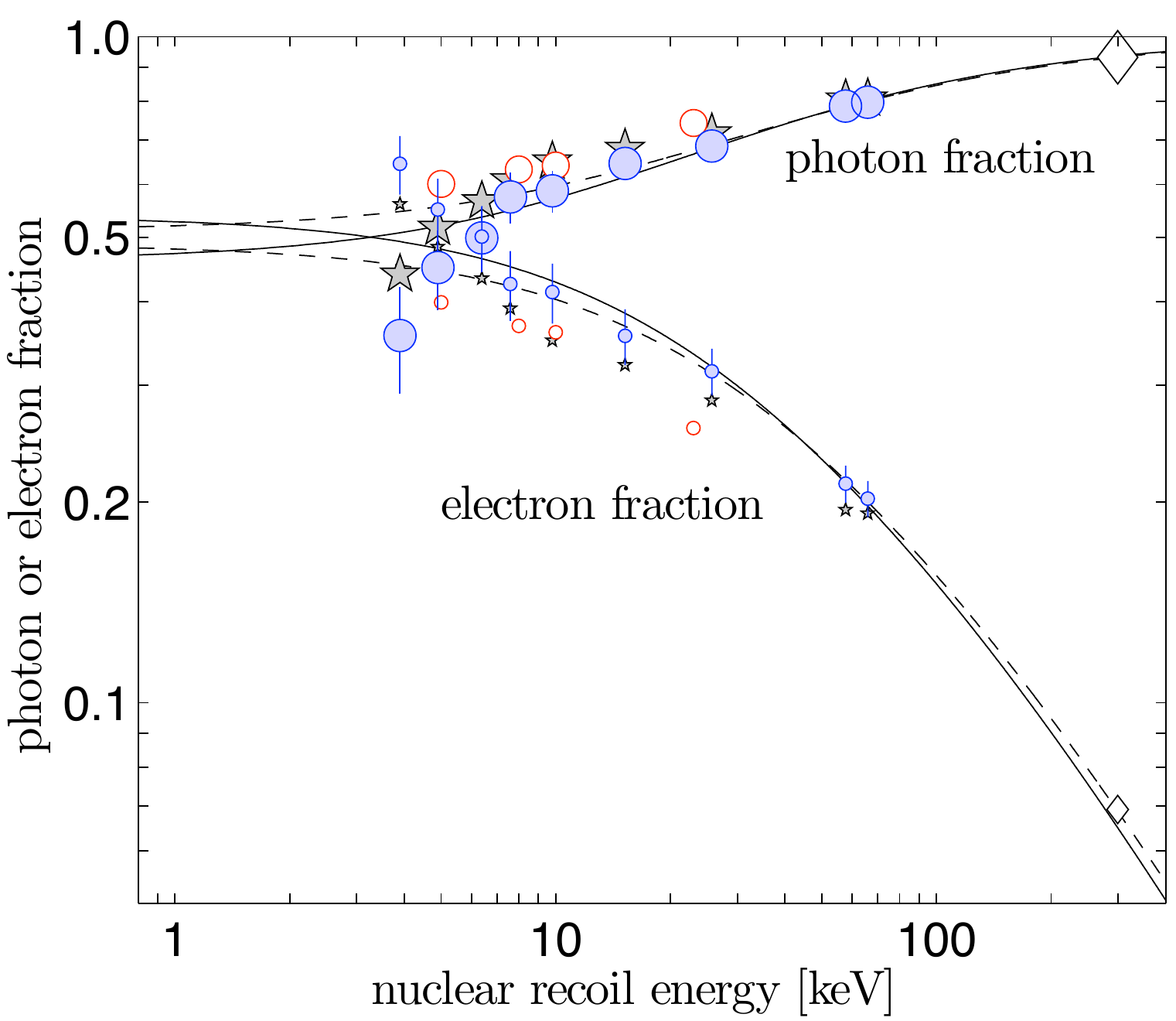}
\caption{The photon fraction $n_{\gamma}/(n_{\gamma}+n_e)$ (large symbols) and the electron fraction $n_e/(n_{\gamma}+n_e)$ (small symbols) for nuclear recoils in liquid xenon, for $E_d=1.0$~kV/cm.  Symbols correspond to data shown in Fig. \ref{fig1}. Uncertainty in $\fullmoon$ and $\bigstar$ were omitted for clarity, and are similar to uncertainty in $\newmoon$.  Curves are a best fit of Eq. \ref{eq6}, as summarized in Table \ref{table1}.}
\label{fig2}
\end{figure}

It may be preferable to use Eq. \ref{eq1a} to reconstruct nuclear recoil energy, instead of E$_{nr}~=~\mbox{S1}/(L_y~\mathcal{L}_{eff}) S_e/S_n$ \cite{2010aprile,2009angle,2008angle}.  In the latter, $S_e$ and $S_n$ are the scintillation reduction due to $E_d$, for electron and nuclear recoils.  In either case, it will be essential to retain a distinct measure of the primary scintillation response.  This is because $\alpha_1\ll\alpha_2$, and so the low-energy signal detection efficiency and energy resolution depend almost entirely on S1.  Additionally, some experiments rely solely on the S1 signal \cite{2009minamino}.  The primary scintillation response can be related to $f_n$ using the photon fraction $n_{\gamma}/(n_{\gamma}+n_e)$, as suggested in \cite{2009manalaysay}.  In Fig. \ref{fig2} we show the photon fraction (large symbols) and the ionized electron fraction $n_e/(n_{\gamma}+n_e)$ (small symbols) for the data from Fig. \ref{fig1}.

It is well-understood that energy loss in liquid noble gases leads to a number $N_{ex}$ of excited atoms and a number $N_i$ of ionized atoms.  We make the usual assumption that each excited xenon atom leads to the creation of one scintillation photon, and that each ionized atom leads to a single electron unless it recombines \cite{2002doke}.  Ionized atoms which recombine result in a single scintillation photon, giving $N_i+N_{ex}=n_e+n_\gamma$ independent of recombination.  The fraction of electrons collected (i.e. those escaping recombination) is predicted by the Thomas-Imel box model \cite{1987thomas} to be
\begin{equation} \label{eq5}
\frac{n_e}{N_i} = \frac{1}{\xi}\mbox{ln}(1+\xi),~~\xi= \frac{N_i\alpha}{4a^2v},
\end{equation}
which describes $N_i$ initial electron-ion pairs in a box of dimension $a$.  This model has been shown to work well for spatially small tracks (nuclear recoils and low-energy electron recoils) when the term $\alpha/(a^2v)$ is held constant. \cite{2009dahl}.  The electron fraction $\mathcal{F}_e = n_e/(n_e+n_\gamma)$ is then given by 
\begin{equation} \label{eq6}
\mathcal{F}_e = \frac{1}{\xi}\mbox{ln}(1+\xi)\frac{N_i}{n_e+n_{\gamma}}, 
\end{equation}
and the photon fraction is $1-\mathcal{F}_e$.  Considering Eq.~\ref{eq1a}, we also have 
\begin{equation}\label{eqNI}
N_i + N_{ex} = \mbox{E}_{nr} f_n / \epsilon.
\end{equation}
We fit Eq. \ref{eq6} to the data, treating $\alpha/(a^2v)$ and $N_{ex}/N_i$ as free parameters and using $f_n$ given by Lindhard with $k=0.166$.  The results are shown in Fig.~\ref{fig2} and summarized in Table \ref{table1}.

\begin{table}[h]
\centering
\caption{Result of fitting Eq. \ref{eq6} to the data shown in Fig. \ref{fig2}.  The first (second) row corresponds to the solid (dashed) curves in Fig. \ref{fig2}.  The XENON10 data point ($\lozenge$) is not included in the fits.}
\begin{tabular}{lccl}
\toprule
data & ~$N_{ex}/N_i$~ & ~$\alpha/(a^2v)$~ & Ref.  \\
\hline
$\newmoon$ & 0.86 & 0.028 & \cite{2010manzur} \\ 
$\newmoon$$^a$ & 1.05 & 0.025 & \cite{2010manzur} \\ 
$\bigstar$ & 1.04 & 0.030 & \cite{2010manzur,2010sorensen}  \\ 
$\fullmoon$ & 1.13 & 0.042 & \cite{2009aprile} \\ 
\botrule
\multicolumn{3}{l}{$^a$ E$_{nr}>7$~keV only }\\
\end{tabular}
\label{table1}
\end{table}

We find that for nuclear recoils in liquid xenon, $N_{ex}/N_i\sim1$ with an uncertainty of about 15\%.  This is in good agreement with \cite{2009dahl}, which found $N_{ex}/N_i=0.89$.  It has been known for some time that for electron recoils, $N_{ex}/N_i=0.06$ \cite{1975takahashi}.  As suggested in \cite{2009dahl}, the roughly factor $\times15$ difference in initial exciton to ion ratios produced by electron and nuclear recoils appears to be the origin of the distinct S2/S1 bands for charged versus neutral particle interactions.  We note that Eq. \ref{eq6} provides a less than satisfactory fit to data ($\newmoon$) with E$_{nr}\lesssim7$~keV, and so we also show the results of a fit which excludes this region.  The disagreement may indicate that $N_{ex}/N_i$ is energy-dependent in this regime.  A detailed discussion of the recombination physics contained in the ratio $\alpha/(a^2v)$ is beyond the scope of this work, and we simply note that our results are very similar to those obtained in \cite{2009dahl}, which used the S2/S1 band centroid rather than fixed-energy data points. 


The observed trend in the electron fraction data ($\newmoon$ and $\bigstar$) implies that by about E$_{nr}\simeq2$~keV, the measured signal may consist almost entirely of electrons.  If the photon fraction is $1-\mathcal{F}_e=0.2$ at 2~keV, and we conservatively assume that Hitachi's calculation of $k=0.110$ is correct, the expected S1 signal at this energy is $\sim3$~photons.  Considering typical values of $\alpha_1$, the probability that this results in a measurable number of photoelectrons is nearly zero.  From the relation
\begin{equation}
\mathcal{L}_{eff} = (1-\mathcal{F}_e)\frac{f_n}{\epsilon} \frac{\alpha_1}{L_y} \frac{S_e}{S_n},
\end{equation}
in which all energy dependence is encoded in $\mathcal{F}_e$ and $f_n$, we would  expect $\mathcal{L}_{eff}=0.025$.  Meanwhile, an electron fraction $\mathcal{F}_e=0.8$ implies an S2 signal of $\sim14$~electrons at 2~keV.  This translates to an ionization yield $\mathcal{Q}_y=6.8$~electrons/keV, or about $\frac{1}{2}\sigma$ higher than the value obtained in \cite{2010sorensen2}.  On the other hand, if the trend predicted by Eq. \ref{eq6} is correct and $1-\mathcal{F}_e=0.5$ at 2~keV, we would expect $\mathcal{L}_{eff}=0.063$.  This implies an S2 signal of 8.5 electrons, and the ionization yield $\mathcal{Q}_y=4.3$~electrons/keV is about $1\sigma$ below the value obtained in \cite{2010sorensen2}.

It has been pointed out that in order for electronic excitation to result from a two-body collision between a projectile (the recoiling atom) and an atomic electron, the maximum possible energy transfer in the collision must exceed the band gap energy E$_g$ of the target \cite{1983ahlen}.  This condition can be expressed in terms of the projectile velocity $v$ and the electron velocity $v_e$ as E$_g<2m_ev(v+v_e)$ \cite{1965gryzinski}.  The electron velocity is often identified with the Fermi velocity \cite{1983ahlen,1987ficenec} $v_F=(3\pi^2\rho_e)^{1/3}\hbar/m_e$.  In the application of the Fermi-gas model to metals, the electron density $\rho_e$ is calculated only for the valence electrons \cite{1976ashcroft}.  This follows from the assumption of a ``free and independent electron gas.''  For insulators or semiconductors, the situation is less clear.  If all 54 electrons in a xenon atom are considered, the requisite projectile velocity which satisfies the inequality (with E$_g=9.3$~eV in xenon \cite{2010aprile3}) implies a cutoff in electronic excitation below a nuclear recoil energy E$_{nr}=39$~keV \cite{2010collar}.  We suggest that it is more physical to calculate $v_F$ under the assumption that the Fermi level E$_F$ lies at the midpoint of the energy gap \cite{1976ashcroft2}.  For xenon this implies an effective 3 valence electrons per atom, with a predicted kinematic cutoff in electronic excitation below E$_{nr}=157$~keV.  Considering Fig. \ref{fig1}, we do not see significant evidence for any kinematic cutoff in excess of the Lindhard prediction.  Applying similar reasoning to germanium, one would consider an effective $1/40$ valence electrons per atom in calculating $v_F$, and predict a kinematic cutoff below E$_{nr}=6.5$~keV.  Again, no evidence for such a cutoff is observed \cite{2007barbeau}.  A simple interpretation of this is that direct two-body collisions are not the dominant electronic energy-loss mechanism for recoiling atoms.  Lindhard {\it et al.} indirectly suggest as much \cite{1963lindhard}.

\section{Summary}
 We have shown that the quenching of electronic excitation from nuclear recoils in liquid xenon  appears well-described by the prediction of Lindhard {\it et al.} \cite{1963lindhard} when taking the appropriate measure of electronic excitation, i.e. the combined energy scale given by Eq. \ref{eq1a}.  Further measurements with reduced uncertainty will be needed to elucidate the correct value of $k$ in Eq. \ref{eq2}, and we have provided three specific suggestions for achieving this.  We have urged the adoption of the combined energy scale for nuclear recoils, with the photon and electron fractions describing the signal partitioning.  This has the distinct advantage of allowing comparison with the quenching theory of Lindhard {\it et al.}, and the recombination model of Thomas and Imel.  

For direct detection experiments with null results, the choice of energy scale has a negligible effect on exclusion limits.  However, one eventually expects to observe nuclear recoils from the scattering of dark matter $-$ optimistically, in the $100+$ live days of blinded data so far accumulated by the XENON100 collaboration.  At that juncture it will be essential to obtain the most accurate reconstruction of the true recoil energy.  We illustrate this point with a simple example, considering the recent XENON100 result (Fig. 3 of \cite{2010aprile}).  Suppose there were two hypothetical nuclear recoil events A and B, both with $\mbox{S1}=4$~photoelectrons, but with event A having log$_{10}(\mbox{S2/S1})=2.3$, and event B having log$_{10}(\mbox{S2/S1})=1.9$.  The S1-only energy scale employed in \cite{2010aprile} assigns the same E$_{nr}=8.7\pm4.3$~keV to both events.  In contrast, the combined energy scale (Eq. \ref{eq1a}) assigns E$_{nr}=8.7\pm2.7$~keV to event A, and E$_{nr}=6.9\pm2.8$~keV to event B, in recognition of the smaller electron signal.  

As shown in Fig. \ref{fig1}, the fraction of energy given to electrons ($f_n$) is expected to (and in fact appears to) decrease with decreasing E$_{nr}$.  But the rate of decrease of $\mathcal{L}_{eff}$ does not necessarily follow the same slope.  This is a consequence of recombination, as shown in Fig. \ref{fig2}.  As of this writing, a direct calibration of the available signal from nuclear recoils with energies $E_{nr}\lesssim4$~keV has remained out of reach.  Yet broad-spectrum neutron calibration data \cite{2010aprile,2008angle}, indicate that liquid xenon direct detection experiments have achieved a sensitivity to nuclear recoils of lower energy.  The question of {\it ``how much lower?''} has necessarily relied on extrapolations of $\mathcal{L}_{eff}$.  In the absence of any theoretical expectation, a reasonable choice has been to assume that $\mathcal{L}_{eff}\simeq0.115$ (flat) for $E_{nr}\lesssim4$~keV \cite{2010aprile}.  We have already shown that this is incompatible with theoretical expectations, if $k=0.110$.  It remains so even if we optimistically assume that Lindhard's calculation of $k=0.166$ is correct.  For example, the expectation at E$_{nr}=2$~keV is then $\mathcal{L}_{eff}=0.036$ for $1-\mathcal{F}_e=0.2$, and $\mathcal{L}_{eff}=0.089$ for $1-\mathcal{F}_e=0.5$.  Given that in all reasonable cases, theoretical expectations suggest that $\mathcal{L}_{eff}$ should decrease with decreasing E$_{nr}$, it is difficult to support the more aggressive of the two exclusion limits (solid curve, Fig. 5 of \cite{2010aprile}) proposed by the XENON100 collaboration.

\section*{Acknowledgments}
This work was partially motivated by an invited talk at the Princeton Center for Theoretical Science workshop ``Dark Matter: Direct Detection and Theoretical Developments,'' Nov 15-16, 2010.  


\end{document}